# A RDF-based Data Integration Framework

Amineh Amini[1], Hadi Saboohi[2], Nasser Nematbakhsh[3]

[1,2] Department of Computer Engineering,
Islamic Azad University, Karaj Branch, Karaj, Iran
{aamini, saboohi} @kiau.ac.ir

[3] Department of Computer Engineering,
Faculty of Engineering, University of Isfahan, Isfahan, Iran
nemat@eng.ui.ac.ir

## Abstract

**Data integration is one of the main problems in distributed data sources. An approach is to provide an integrated mediated schema for various data sources. This research work aims at developing a framework for defining an integrated schema and querying on it. The basic idea is to employ recent standard languages and tools to provide a unified data integration framework. RDF is used for integrated schema descriptions as well as providing a unified view of data. RDQL is used for query reformulation. Furthermore, description logic inference services provide necessary means for satisfiability checking of concepts in integrated schema. The framework has tools to display integrated schema, query on it, and provides enough flexibilities to be used in different application domains.**

## 1. Introduction

Data integration is the problem of combining the data residing at different, heterogeneous sources and providing the user with a unified view of this data, called mediated schema, which can be queried by users. A data integration system frees the user from the knowledge of where and how data are represented in the sources. The interest in this kind of systems has been continuously increased in recent years by the fact that many organizations face the problem of integrating data residing in several different sources. Specifically companies that build data warehouses, doing data mining and developing Enterprise Resource Planning systems must address this problem. Also, integrating data in the World Wide Web is a subject of several investigations and projects nowadays [1]. Using Semantic Web concepts and services for data integration brings new possibilities and challenges. For example adding semantic to the data makes them processable by computer agents. Also it is possible to check for consistency of concepts in the integrated schema.

Early works on integration were carried out in the context of database design, and focused on the so-called schema integration problem, i.e. designing a global, unified schema for a database application starting from several sub schemata, each one produced independently from the others [2]. More recent efforts have been devoted to data integration, which generalizes schema integration by taking into account actual data in the integration process. Here the input is a collection of source data sets (each one constituted by a schema and actual data), and the goal is to provide an integrated and reconciled view of the data residing at the sources, without interfering with their autonomy [3]. We only deal with the so-called read-only integration, which means that such a reconciled view is used for answering queries, and not for updating information.

Data integration can be either virtual or materialized. In the first case, the integration system acts as an interface between the user and the sources [4], and is typical of multi-databases, distributed databases, and more generally open systems. In virtual integration, query answering is generally costly, because it requires accessing the sources. In the second case, the system maintains a replicated view of the data at the sources [5], and is typical, for example, both in information system re-engineering and data warehousing. In materialized data integration, query answering is generally more efficient, because it does not require accessing the sources, whereas maintaining the materialized views is costly, especially when the views must be up-to-date with respect to the updates at the sources (view refreshment). In the rest of this paper, we do not deal with the problem of view refreshment.

There are two basic approaches to the data integration problem, called procedural and declarative. In the procedural approach, data are integrated in an ad-hoc manner with respect to a set of predefined information needs. In this case, the basic issue is to design suitable software modules that access the sources in order to fulfill the predefined information requirements. Several data integration (both virtual and materialized) projects, such as TSIMMIS [6], Squirrel [7], and WHIPS [8] follow this idea. They do not require an explicit notion of integrated data schema, and rely on two kinds of software components: wrappers that encapsulate sources, converting the underlying data objects to a common data model, and mediators that obtain information from one or more wrappers or other mediators, refine this information by integrating and resolving conflicts among the pieces of information from the different sources, and provide the resulting information either to the user or to other mediators. The basic idea is to have one mediator for every query pattern required by the user, and generally there is no constraint on the consistency of the results of different mediators.

In the declarative approach, the goal is to model the data at the sources by means of a suitable language, to construct a unified representation, to refer to such a representation when querying the global information system, and to derive the query answers by means of suitable mechanisms accessing the sources and/or the materialized views. This is the idea underlying systems such as Carnot [9], SIMS [10] and Information Manifold [11]. The declarative approach provides a crucial advantage over the procedural one: although building a unified representation may be costly, it allows maintaining a consistent global view of the information sources, which represents a reusable component of the information integration systems [12].

In this research a declarative approach is adopted and two services namely conceptual modeling of the domain and querying on it are of main concerns. Following technologies are used for that purposes:

1. RDF [13] is used for integrated schema description as well as providing a unified view of data. RDF has a well-defined syntax and data type. Also it has reasonable processing complexity.

2. Description Logic (DL) [14] is used to find any contradiction in the integrated schema (satisfiability of concepts in DL terms). It has well-defined semantic and decidable routines for basic services like satisfiability, which makes it suitable for knowledge representation and reasoning in this domain. We adopt the result of works on DL for databases [15] in this research.

3. RDQL [16] is used for re-formulation of queries. It is a query language for RDF in Jena [17] and provides a data-oriented query model.

The paper is organized as follows. In Section 2, we describe in more detail our framework for data integration based on RDF. In Section 3, implementation of the framework is explained and Section 4 is conclusion.

## 2. Data Integration Framework

Aims of any data integration systems are to build an integrated view of the data defined in various sources and develop a mechanism for data extraction from it. To do so, the framework shall provide following services:

1. For system administrator, it provides facilities to define data sources and integrated schema.

2. Satisfiability checking on integrated schema.

3. For user, displaying integrated schema in various formats (e.g. Entity Relationship model) and mechanism for query execution.

Following a short overview of the framework and its components are described in separate subsections.

In this framework, system administrator defines required data sources as well as integrated schema. Those descriptions are converted to DL statements and are checked for satisfiability. Users can see a visual display of integrated schema and submit their queries. To respond to such queries, system extracts data from various sources, combines them into an integrated data collection, then executes query on it and returns results to the user.

In this framework, wrapper-mediator model [12] for data integration is used. Also in the current implementation, two types of data sources namely database type (relational) and xml type (hierarchical) can be specified.

### 2.1. System Components

Components of the model and their relationships are shown in Fig. 1. An explanation for components follows.

### 2.1.1. Data Source Descriptor File

System administrator defines data sources that are subject of integration. For data sources of database type, URL, username and password together with descriptions of required tables are specified. It is also possible to define a table as a view on database.

If the data source type is "XML Data", local path or URL is specified, and typically an XSL Transformer [18] as a wrapper for converting to the appropriate structure (if needed) is definable.

### 2.1.2. Integrated Schema Descriptor File

Following a relational view of data, integrated schema is defined in the form of (virtual) tables. Each field of such tables can be related to a field in a defined data source. Also, relationships between tables are defined here. These relations can be equality of two fields or two field sets as well as equality relations of one field with a set of fields by equations like mathematical addition or string concatenation.

### 2.1.3. Satisfiability Checker

Integrated schema descriptor file may have contradictory definitions. This module, checks definitions and reports any of such contradictions. In this framework, reasoning procedure as explained in [12] is used.

### 2.1.4. Display Integrated Schema

Integrated schema in Entity-Relationship (ER) diagram and XML formats are displayed. In ER-diagrams, tables, fields and relations between tables are shown graphically. In XML format, in addition field's data types are specified too.

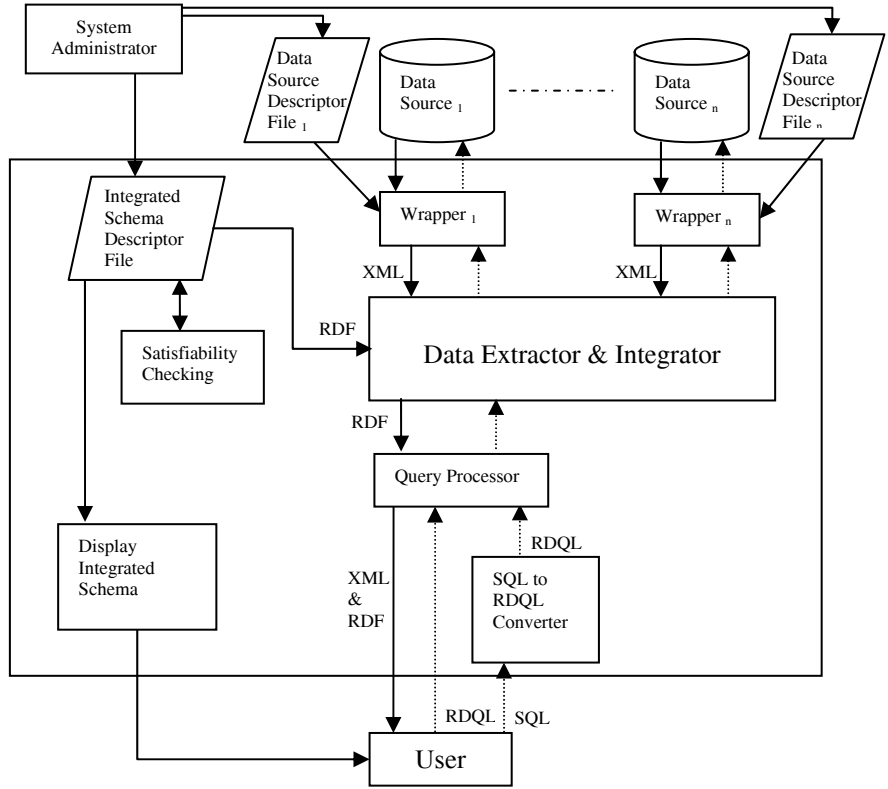

**Fig. 1. Components of the model**

## SQL to RDQL Converter

For query processing, RDQL is used internally. Database users, however, prefer SQL queries. Therefore in this framework, "SQL to RDQL Converter" module is used to convert SQL to RDQL queries. Following algorithm is employed in it:

**SQL to RDQL Converter Algorithm:**
INPUT: a SQL statement over integrated schema.
OUTPUT: a corresponding RDQL statement
METHOD:

1- Create a unique name for any table in SQL query (in "?tableName" form).

2- For the fields in SELECT clause of SQL query, add them to RDQL query with a "?" sign before them.

3- For all the fields appearing in the query, create a property name. A property name is in the "http://integratedDB/integratedTableName#fieldName" format. Having property names, each field is represented as a triple in the WHERE clause of RDQL query. Triples are in "(?tableName propertyName ?fieldName)" format.

4- Finally, conditions in WHERE clause is added to the AND part of RDQL query.

For an example, conversion of following SQL statement to corresponding RDQL statement is demonstrated in Fig. 2:

```
SELECT STUDENT.FIRSTNAME,
       STUDENT.LASTNAME,
       GRADE.AVERAGE,
       STUDENT.DEBT
FROM STUDENT, GRADE
  ON STUDENT.ID=GRADE.STUDENTID
WHERE STUDENT.DEBT>2000
```

---

**Step 1** creates following unique names: ?tbl_0, ?tbl_1
**Step 2** results in: *SELECT ?FIRSTNAME, ?LASTNAME, ?AVERAGE, ?DEBT*
**Step 3** creates following WHERE clause:
*WHERE*
  *(?tbl_0 <http://integratedDB/STUDENT#FIRSTNAME> ?FIRSTNAME),*
  *(?tbl_0 <http://integratedDB/STUDENT#LASTNAME> ?LASTNAME),*
  *(?tbl_1 <http://integratedDB/GRADE#AVERAGE> ?AVERAGE),*
  *(?tbl_0 <http://integratedDB/STUDENT#DEBT> ?DEBT),*
  *(?tbl_0 <http://integratedDB/STUDENT#ID> ?fld_0),*
  *(?tbl_1 <http://integratedDB/GRADE#STUDENTID> ?fld_0)*
**Step 4** adds following phrase to the AND clause of RDQL statement:
*AND ?DEBT > 2000*

**Fig. 2. SQL to RDQL conversion steps**

In the current implementation, SQL statements with aggregation functions, sub queries, and those having expressions in SELECT and ORDER BY clauses are ignored. Those constructs are subject of our future extensions.

### 2.1.5. Query Processor
This module uses Jena API to execute RDQL query on a RDF store that is dynamically created by data extractor and integrator.

### 2.1.6. Data Extractor and Integrator
This module accumulates data from data sources and makes an integrated RDF store from them. The data that only is needed to respond to the user query is accumulated. This is done by considering the table specifications in the query and mapping them, using integrated schema, to actual tables in data sources. Following algorithm is used for this purpose:

**Data Extraction and Integration Algorithm**

1. Read the <Integrated Schema Descriptor (ISD) File> and find all table definitions (ISD tables) and do all following steps for each of them.
2. Make a new table in integrated data and name it with the same name as the name of ISD, call this new integrated data table as <IDT>
3. Read all ISD fields and do steps 4-7 for them
4. Name first ISD field as <master field>, and its data source name as <master source> and its table name as <master table>
5. Scan all <master table> rows in <master source>, and do steps 6 and 7 for them
6. Read the value of all needed fields for ISD in <master table> row, and add them in IDT with appropriate name
7. For each field in ISD that its <source> and <table> is different from <master field>, find a relation between the source and table of it and the source and table of <master field> and search for a row in that table that meets this relation and add the value of requested field to IDT

## 3. Implementation
The system is implemented as a toolkit in Java and provides facilities for defining data sources as well as integrated schema. Queries can be submitted both in SQL and RDQL forms. If query is in SQL, it is first converted to RDQL before submitted to the query processor which executes the RDQL query using Jena. Results can be stored as XML and RDF. A RDF data model can be generated by Jena APIs from the output of "Data extraction and integration algorithm" which also can be subject of user queries. In what follows, main parts of the system are explained.

First step is to create a new "integration project" as shown in Fig. 3.

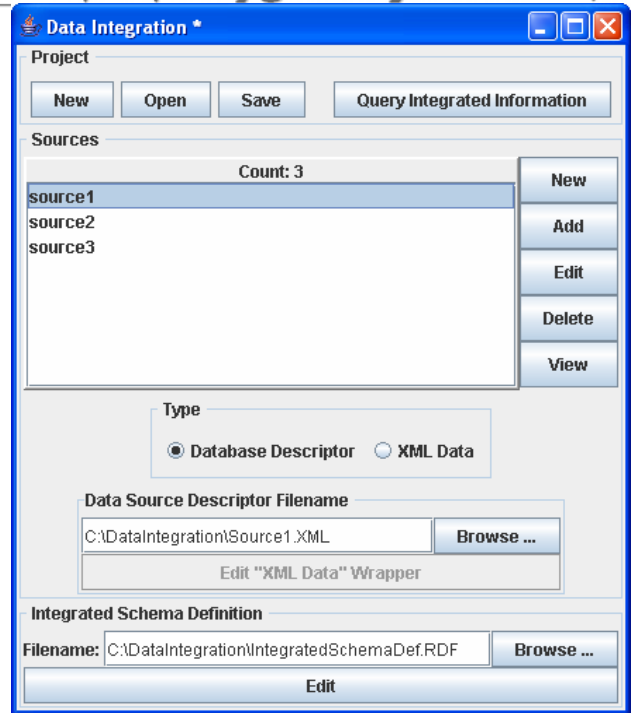

**Fig. 3. Integration system main form**

Data sources should be defined. Currently the system accepts data sources of type database and xml. Also, as shown in the figure, user can specify the integrated schema location and use "Edit" button to make changes to it, if needed.

For database sources, a wrapper description editor form is used as shown in Fig. 4. Following JDBC conventions, this form contains fields for defining connection properties. There are two methods for defining data which are subject to integration. One method is to directly refer to database tables and fields. Alternatively, it is possible to create a virtual view by using a query statement.

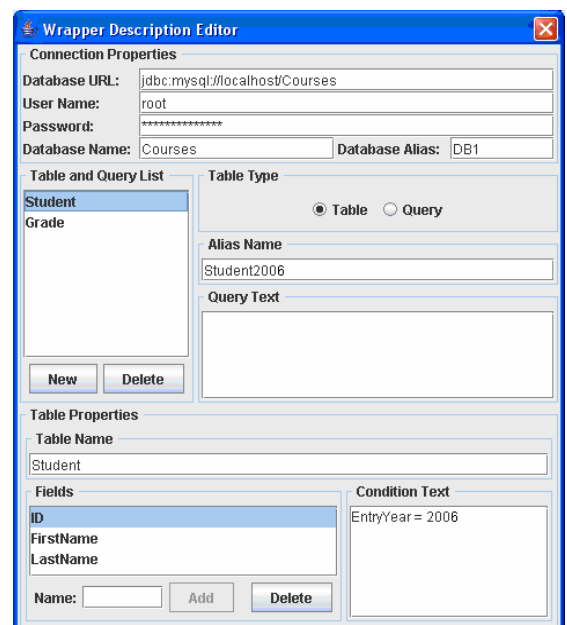

**Fig. 4. Defining a database as a data source**

If the data source is as a "XML Data", these data can be viewed (Read only) as shown in Fig. 5.

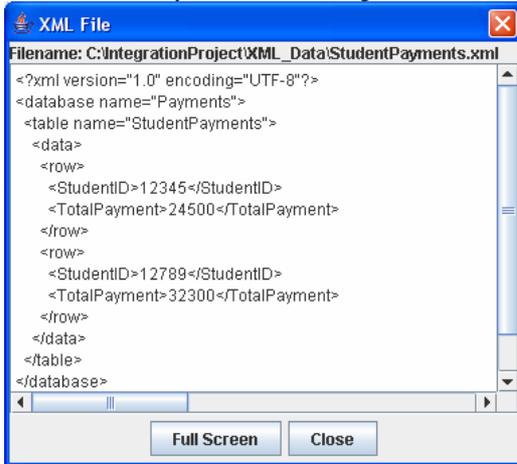

**Fig. 5. Viewing "XML Data"**

After defining data sources, integrated tables and fields should be defined and relations specified. One example is shown in Fig. 6.

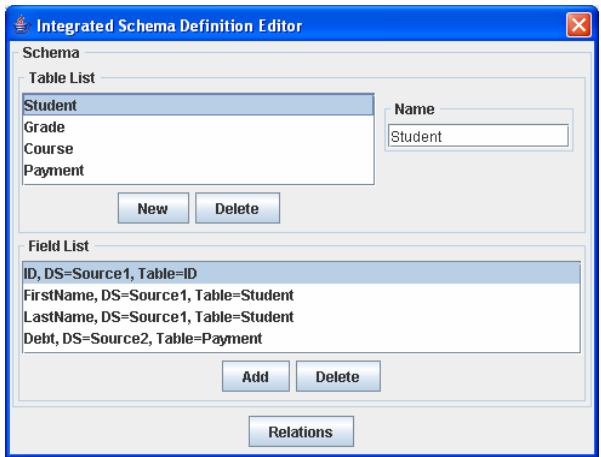

**Fig. 6. Integrated schema definition**

Finally, having all definitions, user can submit queries and see the results in a grid which can be saved as XML and RDF. One example is shown in Fig. 7.

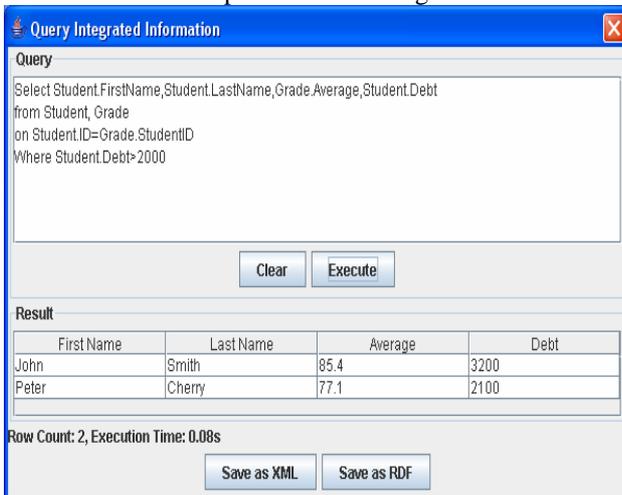

**Fig. 7. Query on integrated data**

In this form, total time for data extraction and integration and query execution is displayed too.

## 4. Conclusion

This paper explained a RDF-based framework for data integration. This framework provides some basic services which can be used for data integration and management of different data sources. These services include satisfiability checking of definitions, displaying data model of integrated schema, query facilities in SQL and RDQL formats and data extraction in XML and RDF formats.

RDF makes it possible to integrate data from different data source types. Also a well-defined query language, RDQL, makes it possible to make queries on such a data model. Jena provides necessary APIs to work on RDF data and execute queries. Therefore we believe the implemented system and idea behind it have many advantages over traditional data integration framework. It is worthwhile to mention that end users (programmers) view the integrated data as like a relational database and can query them using SQL statements, thought, the original sources might not be of type relational data.

One challenge we have now is to make SQL to RDQL converter as general as possible. When such a conversion is fully implemented, there should be no limit on the kinds of queries a user may submit to this system.

Another extension we like to work on is to cache intermediate data extracted from sources. Such intermediate data might be useful for responding to other users' queries. In a real data integration environment this seems to be very useful, considering the fact that data sources might be residing in different distant physical locations.